# A comparison of DA white dwarf temperatures and gravities from *FUSE* Lyman line and ground-based Balmer line observations

M.A. Barstow[1], S.A. Good[1], M.R. Burleigh[1], I. Hubeny[2], J.B. Holberg[3] and A.J. Levan[1]

[1] *Department of Physics and Astronomy, University of Leicester, University Road, Leicester LE1 7RH, UK*

[2] *Laboratory for Astronomy and Solar Physics, NASA/GSFC, Greenbelt, Maryland, MD 20711 USA*

[3] *Lunar and Planetary Laboratory, University of Arizona, Tucson, AZ 85721, USA*

**ABSTRACT**

Observation of the strengths and profiles of the hydrogen Balmer absorption series is an established technique for determining the effective temperature and surface gravity of hot H-rich white dwarf stars. In principle, the Lyman series lines should be equally useful but, lying in the far-UV, are only accessible from space. Nevertheless, there are situations (for example, where the optical white dwarf spectrum is highly contaminated by the presence of a companion) in which use of the Lyman series may be essential. Therefore, it is important to establish whether or not the Lyman lines provide an equally valid means of measurement. We have already made a first attempt to study this problem, comparing Lyman line measurements from a variety of far-UV instruments with ground-based Balmer line studies. Within the measurement uncertainties we found the results from each line series to be broadly in agreement. However, we noted a number of potential systematic effects that could bias either measurement. With the availability of the *FUSE* data archive and observations from our own Guest Observer programmes, we now have an opportunity of examining the use of the Lyman series in more detail from observations of 16 DA white dwarfs. Here we have data produced by a single instrument and processed with a uniform data reduction pipeline, eliminating some of the possible systematic differences between observations of the same or different stars. We have also examined the scatter in values derived from multiple observations of the same star, which is significant. The new results partially reproduce the earlier study, showing that Balmer and Lyman line determined temperatures are in good agreement up to ~50000K. However, above this value there is an increasing systematic difference between the Lyman and Balmer line results, the former yielding the higher temperature. At the moment, there is no clear explanation of this effect but we think that it is most likely associated with deficiencies in the detailed physics incorporated into the stellar model atmosphere calculations. Even so, the data do demonstrate that, for temperatures below 50000K, the Lyman lines give reliable results. Furthermore, for the hotter stars, a useful empirical calibration of the relationship between the Lyman and Balmer measurements has been obtained, that can be applied to other *FUSE* observations.

## 1 INTRODUCTION

In seeking to understand the evolution of white dwarf stars, two of the most important measurements that can be made are of the effective temperature ($T_{\rm eff}$) and surface gravity of any individual object. A key breakthrough in this area was the realization that both $T_{\rm eff}$ and log *g* could be determined from the shape and strength of the profiles of the Balmer absorption lines visible in the optical spectra. This technique was pioneered by Holberg et al.

(1985) and extended to a large sample of white dwarfs by Bergeron, Saffer & Liebert (1992). Combining their results with the theoretical white dwarf evolutionary models of Wood (1992), Bergeron et al. (1992) were able to study the mass distribution of stars in detail. Subsequent studies have taken account of the effects of thin external layers of non-degenerate H and He in the evolutionary models (Wood 1995) and, more recently, the prior evolution of the pre-white dwarf (Blöcker 1995; Driebe et al. 1998).

During the past decade, the Balmer line technique has become the standard method for studying white dwarfs hotter than 10000K to 12000K. Above this temperature range, the Balmer line strengths are monotonic functions of $T_{\rm eff}$ and problems associated with including convection are much diminished. Such work has underpinned the value of various other optical white dwarf surveys (e.g. Bragaglia et al. 1995; Finley, Koester & Basri 1997) and of EUV and X-ray-selected white dwarf samples (Fleming et al. 1996; Marsh et al. 1997; Vennes et al. 1997; Napiwotzki 1999). In these samples, the majority of the stars are isolated objects. If any are in binaries, they are either wide, resolved systems or the companions are late-type dwarfs, where the white dwarf can be spectroscopically isolated. However, when a white dwarf binary companion is spatially unresolved and of type K or earlier, the white dwarf signature is hidden in the glare of the more luminous object and, therefore, the Balmer lines cannot be used for determination of $T_{\rm eff}$ or log *g*. A well-known illustration of this problem is the DA+K star binary V471 Tauri, which has been extensively studied and where the Lyman series spectrum obtained by the *ORFEUS* mission was used to obtain the first accurate measurements of $T_{\rm eff}$ and log *g* (Barstow et al. 1997).

A major result of the EUV sky surveys conducted by *ROSAT* and *EUVE* was the discovery of many unresolved binary systems containing white dwarfs and companion spectral types ranging from A to K (e.g. Barstow et al. 1994b; Burleigh, Barstow & Fleming 1997; Vennes, Christian & Thorstensen 1998). Therefore, a large pool of potential sources exists, for which Lyman series observations are essential to determine $T_{\rm eff}$ and log *g*. Such information can then be coupled with dynamical information from the binary and a *Hipparcos* parallax, measured for the bright companion to provide model-independent estimates of white dwarf mass, testing the evolutionary theories that have been applied to the studies of large isolated samples of white dwarfs.

While the Lyman α line is encompassed by the spectral coverage of *IUE* and *HST*, a single line cannot provide an unambiguous measurement of $T_{\rm eff}$ and log *g*. Access to the full Lyman series lines has been provided by the short duration Hopkins Ultraviolet Telescope (*HUT*) and Orbiting and Retrievable Far and Extreme Ultraviolet Spectrometers (*ORFEUS*) missions. They provided observations of a number of white dwarfs at wavelengths down to the Lyman limit, yielding a first opportunity to compare Balmer and Lyman line measurements systematically. In a previous paper (Barstow et al. 2001), we carried out an evaluation of all the available archival data for these missions, including some early spectra from the Far Ultraviolet Spectroscopic Explorer (*FUSE*). Comparing the results with those from the standard Balmer line analysis, we found general overall good agreement between the two methods. However, significant differences are noted for a number of stars. These differences are not always consistent in that sometimes the Balmer temperature exceeds that derived from the Lyman lines and in other instances is lower. This is not what would be expected if the problems arose from the limitations of the stellar atmosphere calculations and the treatment of the Lyman and Balmer line broadening. We concluded that it is more likely that we were observing systematic effects arising from the observations, the data reduction and the analysis.

In this new paper, we re-examine the issue of the Lyman line analysis with a greatly expanded far-UV data set available from the *FUSE* archive. These spectra cover the complete Lyman line series from β to the series limit, excluding Lyman α. It is now possible to consider a larger number of stars, particularly at values of $T_{\rm eff}$ above 50000K, a range that was sparsely sampled by Barstow et al. (2001). In addition, *FUSE* has observed some of the targets many times, for purposes of monitoring the instrument calibration. Study of the variation in $T_{\rm eff}$ and log *g* within these is a powerful tool for examining systematic effects in the instrument and analysis procedure.

## 2 OBSERVATIONS

### 2.1 Lyman line spectra

All the far-UV spectra for 16 DA white dwarfs utilized in the present paper were obtained by the *FUSE* spectrographs and cover the full Lyman series, with the exception of Lyman α. Table 1 summarizes all the observations, taken by us from the Multi-mission archive (**http://archive.stsci.edu/mast.html**), hosted by the Space Telescope Science Institute. Although flux calibrated data are available in the archive, after processing through the *FUSE* data pipeline, we have reprocessed the data ourselves to provide better control of the quality and a completely uniform calibration, which may vary according to the age of the processed data in the archive. We discuss the data selection and processing below.

The *FUSE* mission was placed in low Earth orbit on 1999 June 24. After several months of in-orbit checkout and calibration activities, science operations began during 1999 December. Therefore, approximately 3 years of data now reside in the archive. An overview of the mission is given by Moos et al. (2000) and its in-orbit performance is described by Sahnow et al. (2000). The spectrograph is described in detail by Green, Wilkinson & Friedman (1994). Further useful information is included in the *FUSE* Observer's Guide (Oegerle et al. 1998), which can be found with other technical documentation on the *FUSE* website (**http://fuse.pha.jhu.edu**).

Although a large number of scientific papers have been published incorporating *FUSE* data, there are several of instrumental issues that affect the quality and usefulness of the spectra in this analysis. Therefore, we give a brief description of the spectrometer and its current status in this context. Based on a Rowland circle design, it comprises four separate optical paths (or channels). Each of these consists of a mirror, a focal plane assembly (including the spectrograph apertures), a diffraction grating and a section of one of two detectors. The channels must be co-aligned so that light from a single target properly illuminates all four channels, to maximize the throughput of the instrument. Two mirrors and two gratings are coated with SiC to provide sensitivity at wavelengths below ~1020Å, while the other two mirror/grating pairs are coated with LiF on Al. This latter combination yields about twice the reflectivity of SiC at wavelengths above 1050Å, but has little reflectivity below 1020Å. The overall wavelength coverage runs from 905 to 1187Å.

**Table 1.** Log of *FUSE* observations of white dwarfs used in this paper. The observation number is a reference allocated in programme ID order for reference within this paper. We also list the aperture size used for each observation (H=HIRS, M=MDRS, L=LWRS)

| Target | obs # | Data ID | Date | Exp time (s) | Ap. | TTAG/ HIST |
|---|---|---|---|---|---|---|
| GD394 | 1 | I8010720000 | 13-Oct-99 | 4940 | L | TTAG |
| | 2 | M1010703000 | 11-Oct-99 | 3502 | L | TTAG |
| | 3 | M1010704000 | 11-Oct-99 | 5652 | L | HIST |
| | 4 | M1010706000 | 13-Oct-99 | 4688 | L | HIST |
| | 5 | M1122201000 | 09-Sep-00 | 4461 | L | TTAG |
| | 6 | P1043601000 | 20-Jun-00 | 28733 | L | TTAG |
| HZ43 | 1 | M1010501000 | 19-Feb-00 | 6092 | L | HIST |
| | 2 | P1042301000 | 22-Apr-00 | 14447 | L | HIST |
| | 3 | P1042302000 | 08-Feb-01 | 39606 | M | HIST |
| | 4 | M1121302000 | 22-Apr-00 | 3671 | L | HIST |
| G191-B2B | 1 | M1010202000 | 17-Feb-00 | 3450 | L | HIST |
| | 2 | M1030403000 | 10-Jan-01 | 483 | H | HIST |
| | 3 | M1030405000 | 25-Jan-01 | 2419 | H | HIST |
| | 4 | M1030506000 | 09-Jan-01 | 503 | M | HIST |
| | 5 | M1030507000 | 10-Jan-01 | 503 | M | HIST |
| | 6 | M1030508000 | 23-Jan-01 | 2418 | M | HIST |
| | 7 | M1030509000 | 25-Jan-01 | 2417 | M | HIST |
| | 8 | M1030604000 | 09-Jan-01 | 503 | L | HIST |
| | 9 | M1030606000 | 23-Jan-01 | 2190 | L | HIST |
| | 10 | M1030607000 | 25-Jan-01 | 1926 | L | HIST |
| | 11 | P1041201000 | 06-Nov-00 | 15451 | H | HIST |
| | 12 | P1041202000 | 13-Jan-00 | 15519 | M | HIST |
| | 13 | S3070101000 | 14-Jan-00 | 15456 | L | HIST |
| GD246 | 1 | M1010601000 | 12-Nov-00 | 1566 | L | HIST |
| | 2 | M1010602000 | 09-Dec-99 | 1236 | M | HIST |
| | 3 | M1010603000 | 09-Dec-99 | 399 | H | HIST |
| | 4 | M1010604000 | 10-Dec-99 | 3371 | L | HIST |
| | 5 | P1044101000 | 19-Jul-00 | 14828 | L | HIST |
| GD153 | 1 | M1010401000 | 06-Mar-00 | 6319 | L | TTAG |
| | 2 | M1010402000 | 29-Apr-00 | 12151 | L | TTAG |
| | 3 | M1010403000 | 07-Feb-01 | 8066 | L | TTAG |
| | 4 | P2041801000 | 28-Jan-01 | 9894 | L | TTAG |
| GD71 | | P2041701000 | 04-Nov-00 | 13928 | L | TTAG |
| GD2 | | P2041101000 | 24-Nov-00 | 10700 | L | TTAG |
| GD659 | 1 | M1010101000 | 04-Jul-00 | 16437 | L | TTAG |
| | 2 | P2042001000 | 11-Dec-00 | 8571 | L | TTAG |
| REJ0457-281 | 1 | P1041101000 | 03-Feb-00 | 19668 | M | TTAG |
| | 2 | P1041102000 | 04-Feb-00 | 10121 | M | TTAG |
| | 3 | P1041103000 | 07-Feb-00 | 17677 | M | TTAG |
| PG1342+444 | | A0340402000 | 11-Jan-00 | 85429 | L | TTAG |
| REJ0558-373 | | A0340701000 | 10-Dec-99 | 11327 | M | TTAG |
| REJ1738+665 | | A0340301000 | 05-May-00 | 6647 | L | TTAG |
| REJ0623-371 | 1 | P1041501000 | 06-Dec-00 | 8371 | L | HIST |
| | 2 | P1041502000 | 03-Feb-01 | 9776 | M | HIST |
| REJ2214-492 | 1 | M1030102000 | 18-Aug-00 | 5796 | H | HIST |
| | 2 | M1030103000 | 24-Oct-00 | 4347 | H | HIST |
| | 3 | M1030201000 | 02-Jun-00 | 4830 | M | HIST |
| | 4 | M1030202000 | 18-Aug-00 | 4830 | M | HIST |
| | 5 | M1030203000 | 24-Oct-00 | 4830 | M | HIST |
| | 6 | M1030305000 | 03-Jun-00 | 5216 | L | HIST |
| | 7 | M1030306000 | 29-Jun-00 | 4193 | L | HIST |
| | 8 | M1030307000 | 17-Aug-00 | 5260 | L | HIST |
| | 9 | M1030308000 | 24-Oct-00 | 4068 | L | HIST |
| | 10 | M1030309000 | 24-Oct-00 | 5083 | L | HIST |
| | 11 | M1030310000 | 24-Oct-00 | 5795 | L | HIST |
| | 12 | M1030311000 | 25-Oct-00 | 6054 | L | HIST |
| | 13 | M1030312000 | 25-Oct-00 | 5491 | L | HIST |
| | 14 | P1043801000 | 03-Jun-00 | 16499 | M | HIST |
| WD1620-391 | | Q1100101000 | 13-Jul-00 | 4830 | M | HIST |
| REJ2334-471 | 1 | M1121702000 | 05-Sep-00 | 3377 | L | HIST |
| | 2 | P1044201000 | 23-Jun-00 | 19356 | L | HIST |
| | 3 | P1044202000 | 07-Nov-99 | 19687 | L | HIST |

**Table 2.** Nominal wavelength ranges (Å) for the *FUSE* detector segments.

| Channel | Segment A | Segment B |
|---|---|---|
| SiC 1 | 1090.9-1003.7 | 992.7-905.0 |
| LiF 1 | 987.1-1082.3 | 1094.0-1187.7 |
| SiC 2 | 916.6-1005.5 | 1016.4-1103.8 |
| LiF 2 | 1181.9-1086.7 | 1075.0-979.2 |

Spectra from the four channels are recorded on two microchannel plate detectors, with a SiC and LiF spectrum on each. Each detector is divided into two functionally independent segments (A and B), separated by a small gap. Consequently, there are eight detector segment/spectrometer channel combinations to be dealt with in reducing the data. The nominal wavelength ranges of these are listed in table 2. Maintaining the co-alignment of individual channels has been difficult in-orbit, probably due to thermal effects. A target may completely miss an aperture for the whole or part of an observation, while being well centred in the others. Hence, in any given observation, not all of the channels may be available in the data. To minimize this problem, most observations have been conducted using the largest aperture available (LWRS, 30 × 30 arcsec). This limits the spectral resolution to between 10000 and 20000, compared to the 24000-30000 expected for the 1.25 × 20 arcsec HIRS aperture. However, this is not important in the analysis presented here, since the Lyman line widths span at least 10Å or more. Various spectra analysed here were obtained through HIRS, MDRS or LWRS apertures and in TIMETAG or HISTOGRAM mode as indicated in table 1.

There are several effects associated with the detectors and electronics, including dead spots and fixed pattern efficiency variations. Potentially, these can lead to spurious absorption features in the processed data if they cut the dispersed spectrum at any point. One particular example is the "worm", which is a strip of decreased flux running along the dispersion direction that can attenuate the incident light by as much as 50%. The "worm" is a shadow cast by the electron repeller grid in the detector, located ~6mm above the MCP surface. In the cross-dispersion direction, the spectrograph optical design places the "vertical" focus at a similar distance above the detector surface. A strong feature is seen if the linear image produced by the grating coincides in 3-dimensional space with one of the grid wires. If the height of the spectral line image above the detector is close to that of the grid wires, the strength of the shadow is very sensitive to the placement of the star in the aperture. In such cases, the flux calibration is not reliable. The shadowing effects are also wavelength dependent and may be manifested as a depression in the flux, for example at ~1100-1140Å in the REJ2334-471 spectrum of Fig. 1. In the spectra we have examined, the "worm" seems to be mostly confined to wavelengths longward of the Lyman lines.

We have used data from both our own Guest Observer programmes and all that available in the public archive upto ~mid 2003 for this work. The targets are confined to hot white DA dwarfs with $T_{\mathrm{eff}}$ above ~20000K. We know that the heavy element content of a white dwarf atmosphere can affect the perceived temperature, determined from the Balmer lines (Barstow, Hubeny & Holberg 1998). Therefore, we have only chosen stars for which we possess enough information about their heavy element content from far-UV observations to assign an appropriate atmosphere composition. Our most recent abundance measurements for these stars can be found in Barstow et al. (2003).

All the data obtained from the archive (table 1) were reprocessed using V2.0.5 of the CALFUSE pipeline. Following the bulk of the analysis reported here, two new versions of the pipeline (V2.2.1, V2.2.3) were released. Therefore, we reprocessed a subset of the data with these latest versions to confirm that any changes made do not affect our results. For example, observation 13 of the G191-B2B series yields $T_{\mathrm{eff}}$=57645K and log $g$=7.60, when processed with CALFUSE V2.0.5 (see table 3). The corresponding values for CALFUSE V2.2.1 are 57702K and 7.58, respectively, in good agreement with the statistical uncertainties. We find a similarly small difference for data processed with versions 2.2.1 and 2.2.3. After reprocessing the data, we considered the separate exposures for a single channel/detector segment. Since the signal-to-noise of these can be relatively poor and the wavelength binning (~0.006Å) over-samples the true resolution by a factor ~10, all the spectra were re-binned to a 0.02Å pixel size for examination. Any strong interstellar absorption features present are used to verify that the wavelength scales for each exposure are well aligned. Individual exposures are then co-added to produce a single spectrum, using our own, specially written IDL routine, weighting the individual spectra by their exposure time. This whole procedure was repeated for all eight detector/channel side combinations.

For the purposes of the analysis of the Lyman lines, we needed to combine the individual spectral segments to provide continuous coverage across the line series. While this is done automatically during the pipeline processing for the archive, there is no mechanism for taking into account variable exposure (from drift in and out of the aperture) or other effects. Therefore, it was necessary for us to develop our own procedure which takes due account of the above problems and which can deal with the differing wavelength ranges and spectral bin sizes of each segment. First, all the spectra are re-sampled onto a common wavelength scale of 0.02Å steps to reduce the level of oversampling. The re-sample/re-binned spectra are then co-added, weighting individual data points by the statistical variance, averaged over a 20Å interval, to take into account any differences in the effective area of each segment and any differences in exposure time that may have arisen from rejection of bad data segments (e.g. those with reduced exposure due to source drift). Through visual inspection, it is apparent that the statistical noise tends to increase towards the edges of a wavelength range. In cases where the signal-to-noise is particularly poor (<3:1 per resolution element) in these regions, we have trimmed the spectra (by ~2-3Å) to remove these data points prior to co-addition. Examples of the resulting spectra, taking one of the higher signal-to-noise data sets for each star, are shown in Fig. 1. The region of poor signal-to-noise seen in the 1080-1087Å region for all the spectra is due to their being no LiF data available at those wavelengths. Where we had multiple exposures for any white dwarf, we carried out a further stage of evaluation of the spectra, discarding those where the signal-to-noise was particularly poor or where the flux levels appeared to be anomalous, when compared to the remainder.

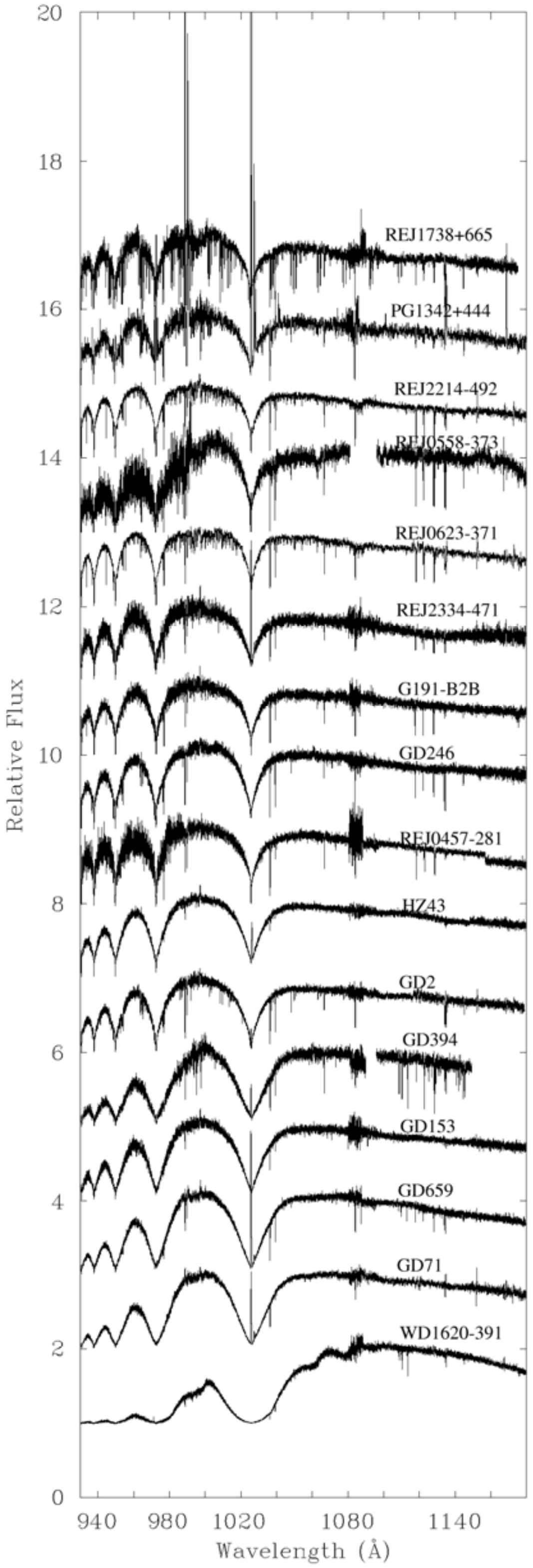

**Figure 1.** Sample *FUSE* spectra for all the DA white dwarfs included in this study, in order of decreasing $T_{\rm eff}$ (as measured with the Balmer lines) from the top of the figure.

### 2.2 Optical spectra

Most of the optical data used here for the Balmer line measurements were obtained as part of a spectroscopic follow-up programme, covering both Northern and Southern hemispheres, for the *ROSAT* X-ray and EUV sky survey. Full details of the observations were originally published by Marsh et al. (1997) and we have made use of these spectra in many subsequent publications (e.g. Barstow, Hubeny & Holberg 1998; Barstow et al. 2001; Barstow et al. 2003). The main difference between the Northern and Southern hemisphere spectra is their resolution, ~8Å (FWHM) and ~3Å (FWHM) respectively. Subsequently, some of the original spectra have been supplemented by more recent observations, where the former did not span the complete Balmer series due to the limited size of the available CCD chip, as discussed by Barstow et al. (2001). For some of the stars we consider here, the "new" observations cover the same wavelength range as the "old" but have better signal-to-noise. We analyse all these data here, as they provide a useful test of the consistency between repeated ground-based observations of the same star, using the same instrument configuration, in the same way as the multiple *FUSE* Lyman line observations discussed above.

## 3 MODEL ATMOSPHERE CALCULATIONS

A key element of our earlier paper (Barstow et al. 2001), comparing Lyman line measurements from *HUT*, *ORFEUS* and early *FUSE* datasets with those from the Balmer lines, is the internal consistency of utilizing a common grid of spectral models. We use the same grids of models here, computed using the NLTE code TLUSTY (Hubeny & Lanz 1995) and its associated spectral synthesis programme SYNSPEC.

The stars included in this study span the complete temperature range of the hot H-rich DA white dwarfs and, therefore, have a range of compositions. The hottest stars contain significant quantities of heavy elements at abundance levels similar to the proto-typical star G191-B2B, while most of the cooler objects have more or less pure-H envelopes (see e.g. Barstow et al. 2003). A few stars have intermediate photospheric compositions. Therefore, we used two appropriate separate grids of models for fitting the Lyman and Balmer lines. For those stars containing heavy elements, which are all the stars with $T_{\rm eff}$ above 50,000 (except for HZ43), we fixed the abundances at the values determined for G191-B2B in an earlier analysis using homogeneous models (He/H = $1.0\times10^{-5}$, C/H = $4.0\times10^{-7}$, N/H = $1.6\times10^{-7}$, O/H = $9.6\times10^{-7}$, Si/H = $3.0\times10^{-7}$, Fe/H = $1.0\times10^{-5}$, Ni/H = $5.0\times10^{-7}$), but taking into account that the CIV lines near 1550Å have subsequently been resolved into multiple components by STIS (Bruhweiler *et al.* 2000). While not all stars have exactly the same heavy element abundances, our recent work has shown that the differences are not very large (Barstow et al. 2003) and, at this level, will not have a significant effect on the Balmer/Lyman line measurements (Barstow et al. 1998). For those stars without significant quantities of heavy elements, we used pure-H models for the analysis.

Since we are concerned with studying the H line profiles in this work, the spectral synthesis code and line broadening included therein are particularly important. In SYNSPEC, we have replaced the hydrogen Stark line broadening tables of Schöning & Butler (private communication) by the more extended tables of Lemke (1997). The latter allow a more accurate interpolation of the electron density for high-density environments, such as the atmospheres of white dwarfs. The spectra produced by the TLUSTY/SYNSPEC codes were recently extensively tested against the results of Koester's codes (Hubeny & Koester in preparation). The differences in the predicted spectra for $T_{\rm eff}$ =60000K and log g=8 were found to be below 0.5% in the whole UV and optical range. Furthermore, we have found that the inaccuracy in the interpolations of the Schöning and Butler tables, together with some fine details of our treatment of the level dissolution, were the primary reason for the disagreement between the spectroscopically deduced $T_{\rm eff}$ using TLUSTY and Koester models obtained by Barstow *et al.* (1998). These changes largely resolve the differences between codes noted by Bohlin (2000).

## 4 DETERMINATION OF TEMPERATURE AND GRAVITY

The technique for determining $T_{\rm eff}$ and log $g$, by comparing the observed line profiles with the predictions of synthetic spectra is well established for the Balmer lines (see Holberg et al. 1986; Bergeron et al. 1992 and many other subsequent authors). We have described our own Balmer line analysis technique in earlier papers (e.g. Barstow et al. 1994a). The Lyman line analysis technique is similar, but we have developed the exact approach during a series of papers and, therefore, reiterate the current procedures here.

Analysis of both series of lines was performed using the program XSPEC (Shafer et al. 1991), which adopts a $\chi^2$ minimization technique to determine the model spectrum giving the best agreement with the data. For the Balmer lines, we include Hβ through to Hε simultaneously in the fit. Applying an independent normalization constant to each ensures that the result is independent of the local slope of the continuum, reducing the effect of any systematic errors in the flux calibration of the spectrum.

In contrast to the ground-based Balmer line series observations, the Lyman series are obtained using space-based platforms. Hence, there are no systematic errors arising from an atmospheric extinction correction. Furthermore, the flux calibration is usually obtained from a detailed off-line calibration of the instrument, applied as part of the standard pipeline, rather than direct comparison of the spectrum with that of a selected standard star. Since the instrument calibration still refers to observations of well-studied stars, such as white dwarfs, some systematic uncertainties will still apply, but will be different from and independent of those arising from ground-based techniques.

Unlike the Balmer lines, which can be isolated individually (Fig. 2), the Lyman series generally overlap substantially shortward of Lyman β. Therefore, we divided the *FUSE* data into two wavelength ranges: from 1000Å to 1050Å, incorporating the β line, and from 930Å to 990Å, covering the remaining lines (γ through ε inclusive). To take account of any low frequency systematic effects in the flux calibration, we applied individual normalization constants to each of the two sections of data.

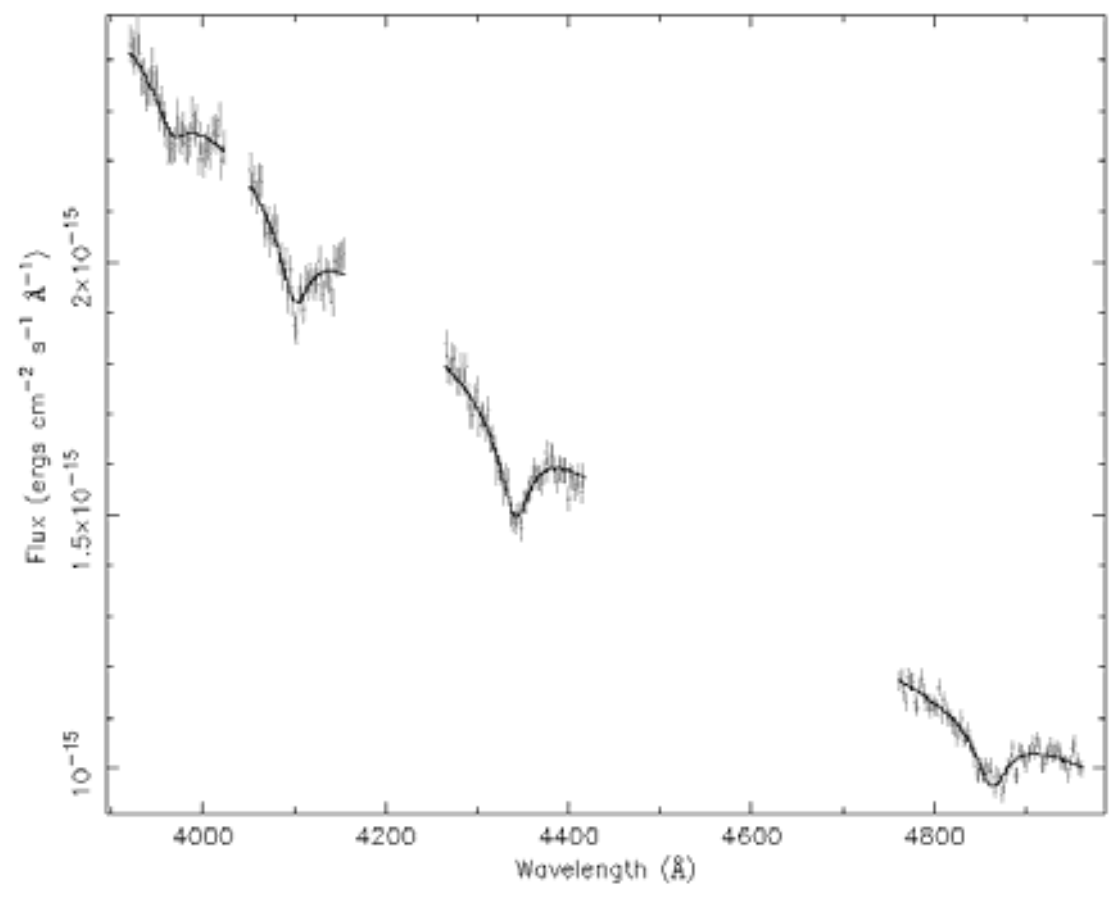

**Figure 2.** Model atmosphere fit (black histogram) to the Balmer line spectrum of PG1342+444 (grey error bars).

Apart from applying an extinction correction and flux calibration, analysis of the Balmer lines is straightforward (Fig. 2), since the lines are, in general, uncontaminated by any other components, unless the star has a binary companion and the spectrum composite. However, many of the Lyman series spectra show strong emission lines, due to the Earth's geocorona, superimposed on the white dwarf spectrum. Secondly, interstellar HI absorption can artificially deepen the core of the Lyman absorption lines. These two effects compete with each other and occasionally conspire to cancel each other out, but usually they must be removed from the data in an appropriate way.

Observed geocoronal line strengths are determined primarily by intensity of the resonantly scattered solar Lyman emission lines, temporal factors and observing geometry. In general, Lyman β is the most intense with the higher order lines showing a strong progressive decrease. However, even if no obvious emission feature seen in any Lyman absorption line, there may still be a contribution which distorts the line core from its natural shape. The geocoronal radiation has a natural Doppler width, which is further broadened in the instrument by the diffuse nature of the source and the spectral resolution. For the LWRS, for example, the airglow lines will have a width ~0.3Å and weak emission features may distort the stellar spectrum without being immediately obvious. As a result of differences in the relative velocity of the stellar and geocoronal sources, the emission lines are not necessarily aligned with the stellar cores. In this work we ignore the central regions of each Lyman line core to remove the contamination from both geocoronal and interstellar material, as illustrated by the data gaps in Fig. 3. We also remove all other interstellar absorption lines.

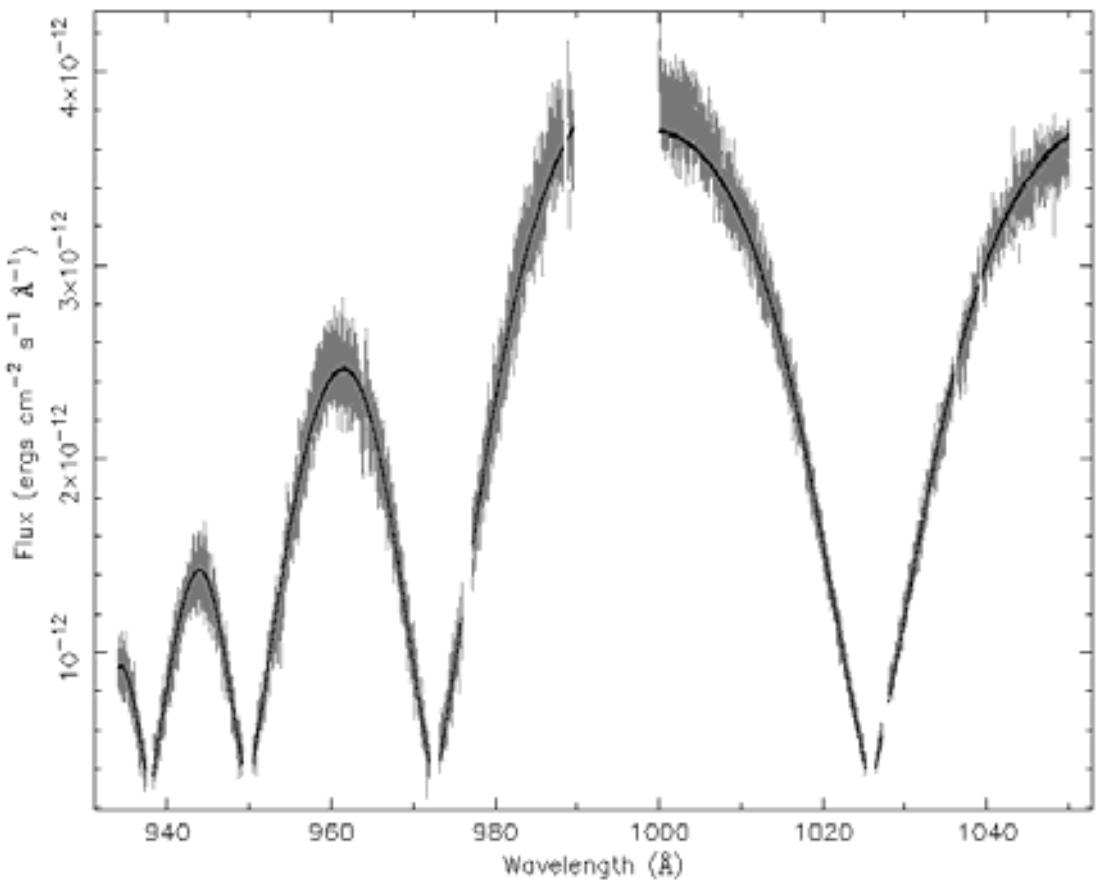

**Figure 3.** Lyman β-ε lines from a *FUSE* spectrum of the hot DA white dwarf GD659, showing the removal of the Lyman line cores to avoid contamination from geocoronal emission and interstellar absorption. Strong interstellar lines have also been removed for the analysis. The best-fit spectrum is represented by the black curve and the data by the grey error bars.

The white dwarf GD659, shown in Fig. 3, has an atmosphere largely devoid of heavy elements. Although C, N and Si are detected by *HST*, at longer wavelengths than covered by *FUSE*, no features (other than interstellar lines) are visible above the signal-to-noise of the data considered here. However, this is not true for all white dwarfs in the sample. In particular, numbers of heavy element absorption lines are visible in the spectra of the hottest, heavy element-rich stars, as shown very clearly in Fig. 1. In principle, the heavy element lines can be treated explicitly in the model calculations and synthetic spectra. However, there are two problems with this approach. The *FUSE* spectral range is only now becoming well studied. In the context of white dwarf observations, it is not certain that we have taken into account all the possible transitions in the models and we have little experience regarding the reliability of the atomic data used. Furthermore, as we have seen in the STIS spectra, the detailed line profiles can be strongly affected by the stratification of atmospheric material (see Barstow et al. 2003), an effect not included in the present calculations. All the evidence we have from our recent work (e.g. Barstow et al. 1998) suggests that these details only have a secondary influence on the shape and strength of either the Lyman or Balmer series lines. However, from a statistical point of view, strong heavy element absorption lines could have an undue influence on the overall fitting procedure, particularly if they are not accurately reproduced by the assumed value and depth dependence (homogeneous in this case) of the abundance. Consequently, we have removed all significant photospheric lines from the spectra during the analysis, as shown for REJ1738+665 (Fig. 4). Comparing this spectrum with that of GD659 in Fig. 3, it can be seen that there is an increased number of gaps in the data, from where the heavy element lines have been excised. We note that for REJ1738+665, we have also had to remove a number of interstellar $H_2$ lines that are present in the spectrum of this object.

We have analysed *FUSE* and ground-based data for a total of 16 hot DA white dwarfs with temperatures ranging from 67000K (REJ1738+665) down to 22500K (WD1620-391), determined from the Balmer lines. These results are listed in table 3. The 1σ errors listed are the formal statistical uncertainties determined by allowing $T_{\rm eff}$ and log $g$ to vary until $\delta\chi^2$ reaches a value of 2.3, corresponding to the 1σ level for two degrees of freedom. It should be noted that these errors do not take into account possible, significant, contributions from systematic effects related to the data reduction and calibration processes. As we have discussed in our earlier paper on Lyman series studies (Barstow et al. 2001), such effects can often be a factor 2-3 larger than the statistical values. Having repeated observations of several stars with the same instrument affords an opportunity to examine this particular problem in detail. Consequently, table 3 lists separately all the values of $T_{\rm eff}$ and log $g$ obtained for those stars with multiple observations.

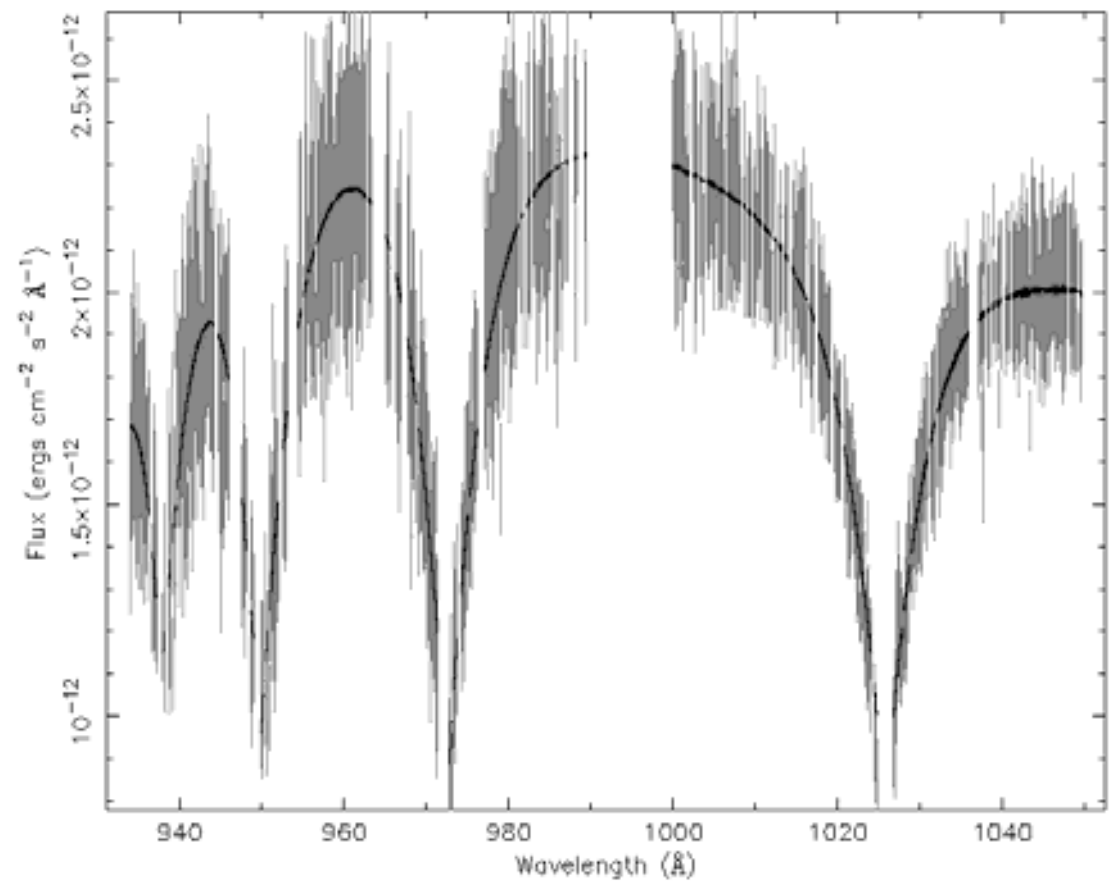

**Figure 4.** Lyman β-ε lines from a *FUSE* spectrum of the hot DA white dwarf REJ1738+665, showing the removal of the Lyman line cores to avoid contamination from geocoronal emission and interstellar absorption. Strong interstellar (including $H_2$) and photospheric lines have also been removed for the analysis. The best-fit spectrum is represented by the black curve and the data by the grey error bars.

**Table 3.** Values of $T_{eff}$ and log *g* obtained for all far-UV and ground-based optical observations of the 16 white dwarfs included in this study. All the 1σ errors quoted for each spectrum are the statistical uncertainties, while the 1σ errors listed for the simple means of the multiple observations are derived from the sample variance for each individual star. Observation numbers for the Lyman series measurements refer to the spectra listed in table 1.

| | Balmer line measurements | | | | | Lyman line measurements | | | | Mean Balmer line values | | | | Mean Lyman line values | | | |
|---|---|---|---|---|---|---|---|---|---|---|---|---|---|---|---|---|---|
| Star | $T_{eff}$ (K) | error | log *g* | error | No | $T_{eff}$ (K) | error | log *g* | error | $T_{eff}$ (K) | error | log *g* | error | $T_{eff}$ (K) | error | log *g* | error |
| REJ1738+665 | 66760 | 1230 | 7.77 | 0.10 | | 75799 | 685 | 7.850 | 0.033 | | | | | | | | |
| PG1342+444 | 66750 | 2450 | 7.93 | 0.11 | | 54308 | 869 | 7.901 | 0.075 | | | | | | | | |
| REJ2214-492 | 61600 | 2300 | 7.29 | 0.11 | 1 | 70657 | 265 | 7.590 | 0.025 | | | | | 70769 | 1297 | 7.590 | 0.068 |
| | | | | | 2 | 70884 | 270 | 7.538 | 0.026 | | | | | | | | |
| | | | | | 3 | 72109 | 243 | 7.658 | 0.021 | | | | | | | | |
| | | | | | 4 | 70583 | 190 | 7.737 | 0.018 | | | | | | | | |
| | | | | | 5 | 72813 | 206 | 7.560 | 0.018 | | | | | | | | |
| | | | | | 6 | 71948 | 149 | 7.537 | 0.014 | | | | | | | | |
| | | | | | 7 | 70243 | 150 | 7.553 | 0.015 | | | | | | | | |
| | | | | | 8 | 70920 | 140 | 7.570 | 0.014 | | | | | | | | |
| | | | | | 9 | 68247 | 173 | 7.610 | 0.015 | | | | | | | | |
| | | | | | 10 | 70994 | 144 | 7.562 | 0.014 | | | | | | | | |
| | | | | | 11 | 68396 | 161 | 7.535 | 0.015 | | | | | | | | |
| | | | | | 12 | 72037 | 147 | 7.534 | 0.019 | | | | | | | | |
| | | | | | 13 | 70061 | 480 | 7.553 | 0.043 | | | | | | | | |
| | | | | | 14 | 70873 | 109 | 7.721 | 0.010 | | | | | | | | |
| REJ0558-373 | 59510 | 2210 | 7.70 | 0.14 | | 65862 | 598 | 7.610 | 0.085 | | | | | | | | |
| REJ0623-371 | 58200 | 1800 | 7.14 | 0.11 | 1 | 64711 | 172 | 7.546 | 0.016 | | | | | 65757 | 1479 | 7.540 | 0.004 |
| | | | | | 2 | 66803 | 215 | 7.541 | 0.020 | | | | | | | | |
| REJ2334-471 | 53205 | 1300 | 7.67 | 0.10 | 1 | 56196 | 410 | 7.704 | 0.034 | | | | | 55933 | 510 | 7.720 | 0.020 |
| | | | | | 2 | 55345 | 128 | 7.746 | 0.007 | | | | | | | | |
| | | | | | 3 | 56258 | 250 | 7.705 | 0.018 | | | | | | | | |
| G191-B2B | 51510 | 880 | 7.53 | 0.09 | 1 | 58706 | 150 | 7.654 | 0.011 | 53035 | 2156 | 7.57 | 0.05 | 58152 | 831 | 7.640 | 0.061 |
| | 54560 | 200 | 7.60 | 0.02 | 2 | 56950 | 185 | 7.573 | 0.050 | | | | | | | | |
| | | | | | 3 | 57834 | 308 | 7.608 | 0.022 | | | | | | | | |
| | | | | | 4 | 58352 | 43 | 7.715 | 0.036 | | | | | | | | |
| | | | | | 5 | 58388 | 53 | 7.706 | 0.036 | | | | | | | | |
| | | | | | 6 | 58687 | 233 | 7.712 | 0.015 | | | | | | | | |
| | | | | | 7 | 59183 | 278 | 7.720 | 0.018 | | | | | | | | |
| | | | | | 8 | 58690 | 413 | 7.691 | 0.030 | | | | | | | | |
| | | | | | 9 | 58141 | 360 | 7.625 | 0.030 | | | | | | | | |
| | | | | | 10 | 57098 | 175 | 7.633 | 0.010 | | | | | | | | |
| | | | | | 11 | 56859 | 181 | 7.541 | 0.015 | | | | | | | | |
| | | | | | 12 | 57645 | 110 | 7.600 | 0.005 | | | | | | | | |
| | | | | | 13 | 59441 | 78 | 7.589 | 0.060 | | | | | | | | |
| GD246 | 51300 | 850 | 7.91 | 0.07 | 1 | 52289 | 318 | 7.843 | 0.030 | | | | | 52402 | 868 | 7.890 | 0.090 |
| | | | | | 2 | 51396 | 430 | 8.000 | 0.020 | | | | | | | | |
| | | | | | 4 | 53514 | 246 | 7.782 | 0.021 | | | | | | | | |
| | | | | | 5 | 52411 | 219 | 7.879 | 0.020 | | | | | | | | |
| REJ0457-281 | 50960 | 1070 | 7.93 | 0.08 | 1 | 67020 | 207 | 8.000 | 0.003 | | | | | 66821 | 282 | 8.000 | 0.010 |
| | | | | | 3 | 66621 | 265 | 8.000 | 0.002 | | | | | | | | |
| HZ43 | 50370 | 780 | 7.85 | 0.07 | 1 | 50203 | 90 | 7.970 | 0.013 | 49435 | 1322 | 7.95 | 0.14 | 50377 | 324 | 7.970 | 0.030 |
| | 48500 | 400 | 8.05 | 0.02 | 2 | 50178 | 92 | 7.945 | 0.009 | | | | | | | | |
| | | | | | 3 | 50751 | 40 | 8.000 | 0.003 | | | | | | | | |
| GD2 | 45460 | 510 | 7.87 | 0.06 | | 46196 | 138 | 7.715 | 0.017 | | | | | | | | |
| GD394 | 39290 | 360 | 7.89 | 0.05 | 1 | 35803 | 105 | 8.194 | 0.042 | | | | | 35851 | 233 | 8.320 | 0.080 |
| | | | | | 2 | 35480 | 75 | 8.251 | 0.029 | | | | | | | | |

**Table 3. continued**

| | Balmer line measurements | | | | | Lyman line measurements | | | | Mean Balmer line values | | | | Mean Lyman line values | | | |
|---|---|---|---|---|---|---|---|---|---|---|---|---|---|---|---|---|---|
| Star | $T_{eff}$ (K) | error | log *g* | error | No | $T_{eff}$ (K) | error | log *g* | error | $T_{eff}$ (K) | error | log *g* | error | $T_{eff}$ (K) | error | log *g* | Error |
| GD 394 | | | | | 3 | 36028 | 65 | 8.399 | 0.021 | | | | | | | | |
| | | | | | 4 | 36164 | 121 | 8.394 | 0.024 | | | | | | | | |
| | | | | | 5 | 35819 | 70 | 8.332 | 0.026 | | | | | | | | |
| | | | | | 6 | 35815 | 21 | 8.312 | 0.008 | | | | | | | | |
| GD153 | 39290 | 340 | 7.77 | 0.05 | 1 | 38745 | 92 | 7.861 | 0.019 | 38205 | 1534 | 7.9 | 0.18 | 38487 | 247 | 7.870 | 0.010 |
| | 37120 | 140 | 8.02 | 0.02 | 2 | 38578 | 67 | 7.875 | 0.019 | | | | | | | | |
| | | | | | 3 | 38463 | 67 | 7.880 | 0.020 | | | | | | | | |
| | | | | | 4 | 38160 | 64 | 7.882 | 0.015 | | | | | | | | |
| GD659 | 35660 | 135 | 7.93 | 0.03 | 1 | 34848 | 39 | 7.935 | 0.021 | | | | | 34704 | 204 | 7.910 | 0.03 |
| | | | | | 2 | 34560 | 53 | 7.887 | 0.016 | | | | | | | | |
| GD71 | 32780 | 65 | 7.83 | 0.02 | 3 | 32048 | 30 | 7.833 | 0.011 | | | | | | | | |
| WD1620-391 | 22450 | 45 | 8.22 | 0.01 | | 22501 | 10 | 7.501 | 0.002 | | | | | | | | |

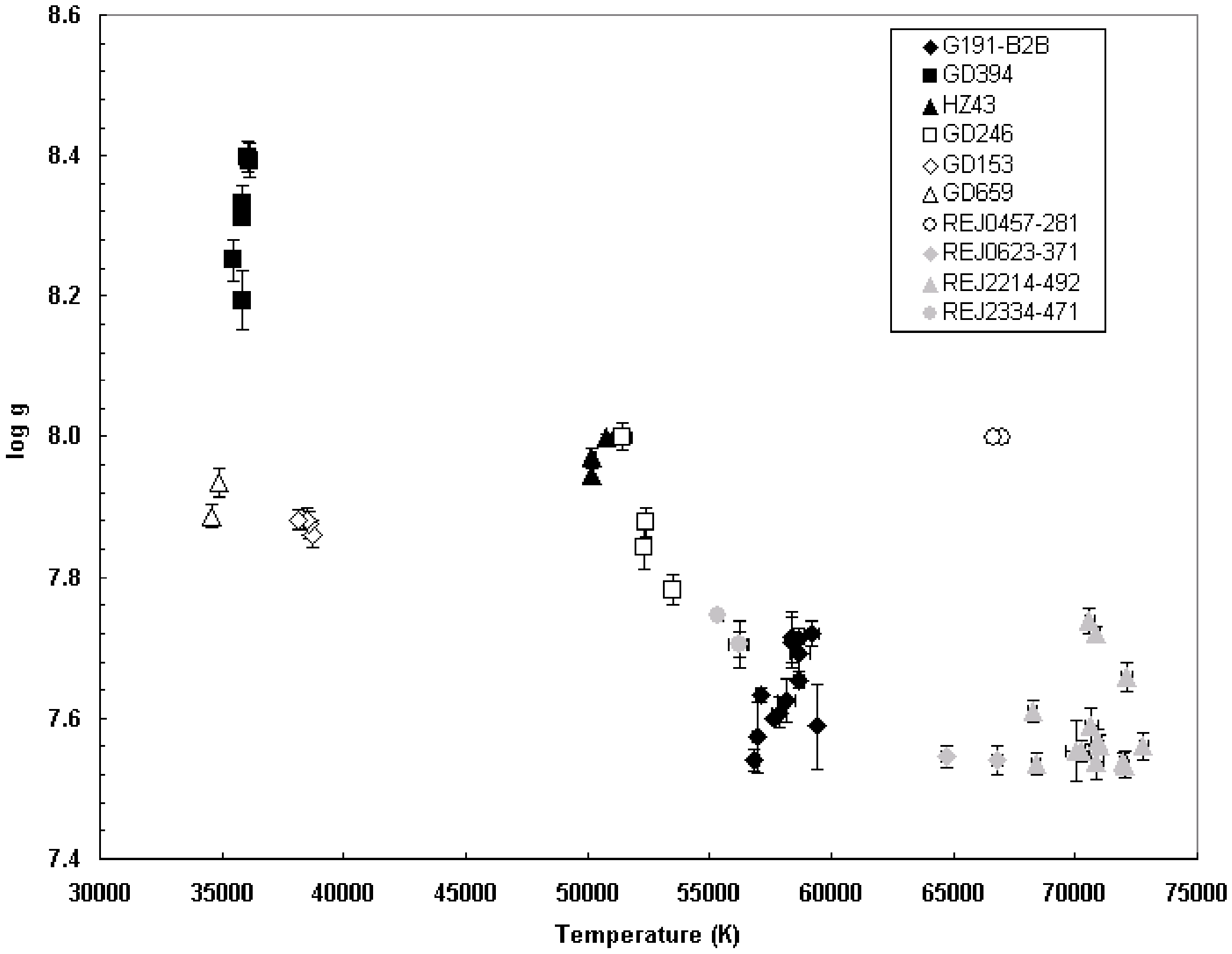

**Figure 5.** The Lyman line determinations of $T_{eff}$ and log *g* for all multiple white dwarf observations. The individual white dwarfs in question are indicated by the symbols listed in the key. The error bars shown are the statistical 1σ uncertainties. Where these are not visible, the error is smaller than the size of the plot symbol.

## 5 DISCUSSION

### 5.1 Multiple observations of individual stars

White dwarfs are often used as photometric calibration sources for UV instruments due to their flux stability and relative ease of modeling the flux distribution. Consequently, many of the targets included in this work have been observed more than once, as shown in table 3. Indeed, 10 of the 16 white dwarfs studied have been observed more than once and some many times. For example, we have 13 separate observations of G191-B2B and 14 for REJ2214-492. It is clear, merely from inspection of table 3 that there is some scatter in values of both $T_{eff}$ and log *g* obtained for individual stars. To appreciate the range of this scatter in the context of the statistical uncertainties and the spread of values for the sample as a whole, all the individual Lyman line measurements are displayed in the $T_{eff}$/log $g$ plane in Fig. 5. It can be seen that the scatter in data values is larger than the typical systematic errors applied to both $T_{eff}$ and log $g$. In general, the extent of the variation is related to the number of observations in each white dwarf sample. This might indicate some standard distribution, such as a gaussian, but the number of measurements is too small to prove this for any of the stars studied.

Whatever the particular reason for the spread in values of $T_{eff}$ and log $g$, the observed scatter is an important indication of the true measurement error. If we make the assumption that the systematic effects produce a normal distribution of uncertainty, then we can calculate a simple mean value and assign a 1σ uncertainty from the variance. The results of this exercise are summarised in Table 3, for both Balmer and Lyman line analyses. This reduction of repeated measurements to a single value is most appropriate where the sample size is large. At most we only have two Balmer line observations for each star. For the Lyman lines, we have a wider range of multiple values, from 2 to as many as 14, for REJ2214-492.

The large number of values we have for both REJ2214-492 and G191-B2B (14 and 13 respectively) gives the best comparison of the total measurement error with the formal statistical value. For example, the 1σ statistical errors on $T_{eff}$ range from 43K to 360K for G191-B2B, compared to 831K computed from the sample variance. Similarly the statistical range of 109K to 480K for REJ2214-492 is smaller than the 1297K determined from the sample scatter. In our previous paper on Lyman line measurements (Barstow et al. 2001), we suggested (as also indicated by other authors) that the statistical errors underestimated the true uncertainties by a factor 2-3. This is so for the largest statistical errors, determined for the spectra with the lowest signal-to-noise. However, when the signal-to-noise is increased the underestimation factor is much larger (10-20) for these stars.

Figs. 6 and 7 show how the statistical errors in $T_{eff}$ and log $g$ depend upon the signal to noise of each observation of REJ2214-492, which we have computed from the measured average flux in the 1050-1055Å range and its associated statistical error. There is a clear relationship with smaller errors associated with increased signal-to-noise and a one-to-one correspondence between the results for $T_{eff}$ and log $g$, albeit with some scatter. However, the best information we have on the overall uncertainties comes from the scale of the sample variance, which is ~1.5% of $T_{eff}$ for G191-B2B and REJ2214-492. The errors in log $g$ are of a similar magnitude. It is important to note that there is no correlation between the magnitude of the statistical errors (and by implication the signal-to-noise) and the measured values of $T_{eff}$ or log $g$.

Both REJ2214-491 and G191-B2B are among the hottest stars in the group of DA white dwarfs we have studied. Unfortunately, no cooler DAs have been observed so intensively. However, the observations of GD394 (6) and GD153 (4) suggest that that the systematic measurement uncertainties decrease by no more than a factor ~2 toward the cooler end of the temperature range.

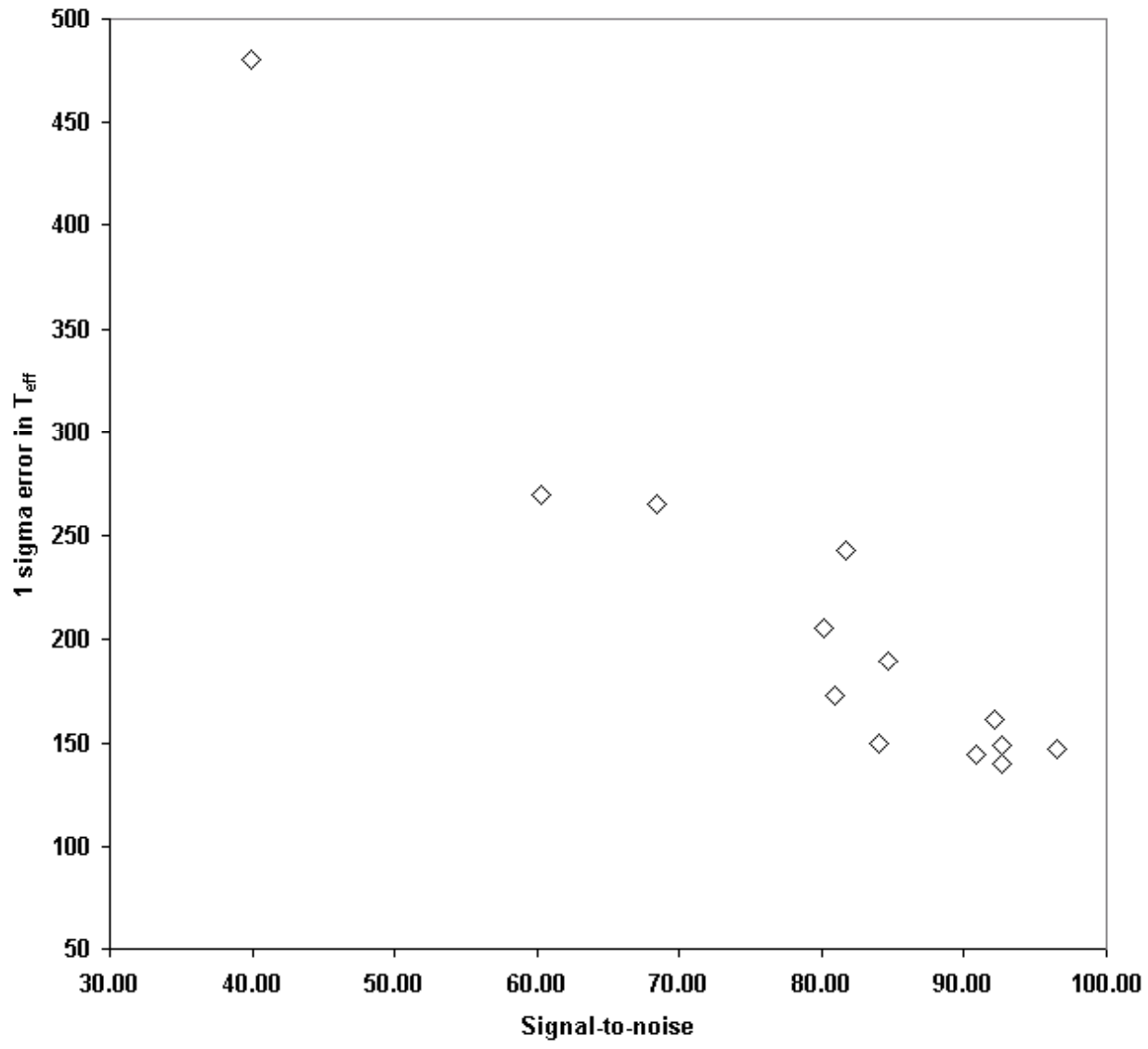

**Figure 6.** The 1σ statistical error on $T_{eff}$ measurements for the individual observations of REJ2214-492, as a function of the average signal-to-noise in the 1050-1055Å range.

It is clear from this analysis that, however much we improve the signal-to-noise of the observations, effects other than the simple photon counting statistics ultimately limit us. So far we have not considered the possible sources of the measurement scatter that we see and to decide on which are the most important will probably require a controlled series of observations. Nevertheless we can eliminate some possibilities and discuss the magnitude of others. Since we are using the same instrument, pipeline processing/calibration and stellar atmosphere models for the spectral analyses, we can already eliminate these factors. Hence, we need only consider the differences that can arise between specific observations. One possible variable is the level of the background, which can vary with time and geometrical factors associated with location of the satellite in its orbit (e.g. day/night conditions) and with respect to the earth. These will rarely be repeated identically from one observation to the next. Furthermore, Barstow et al. (2001) have already shown that errors of a few percent in subtraction of background or scattered light components can give rise to errors of the size noted here.

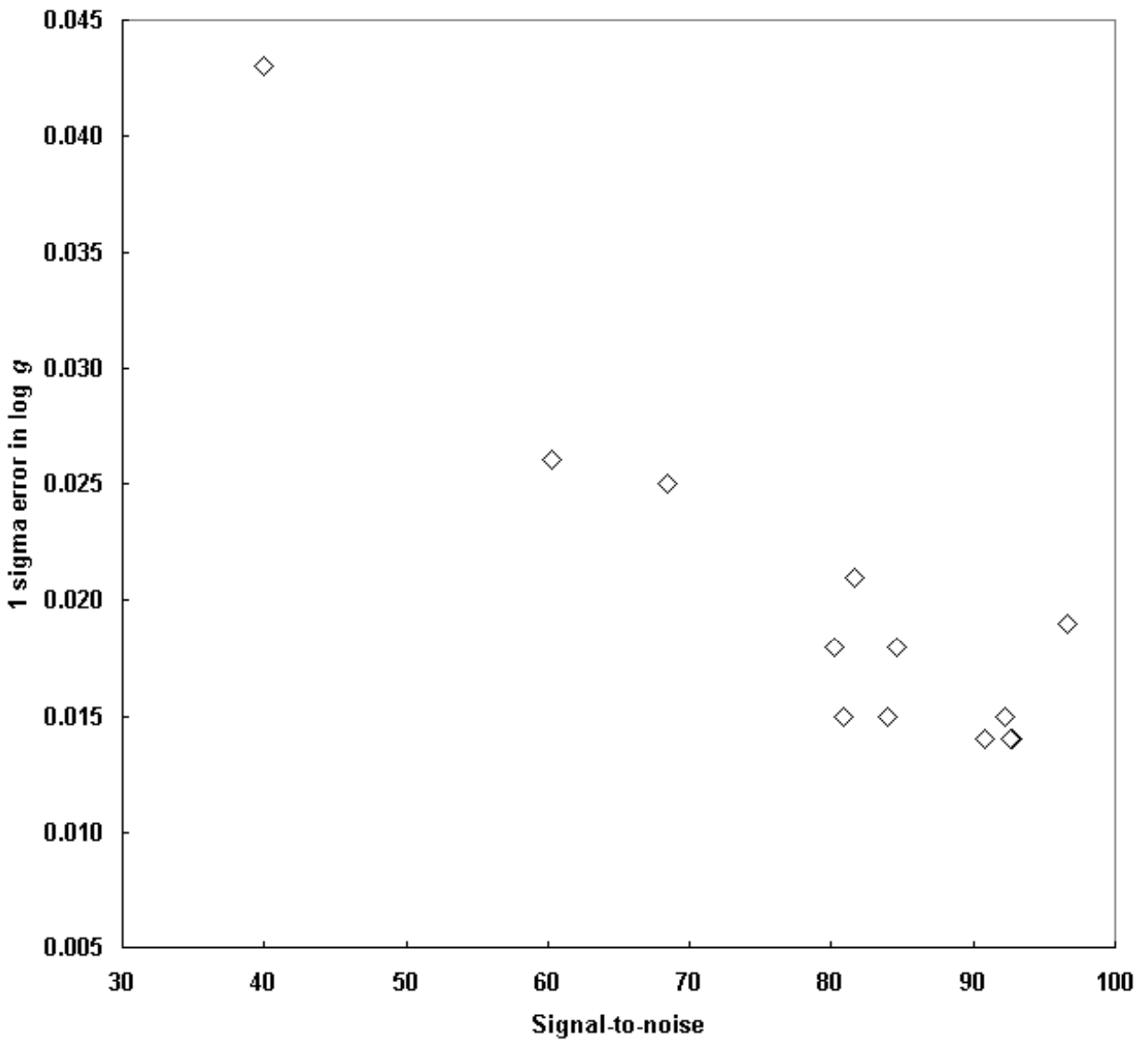

**Figure 7.** The 1σ statistical error on log *g* measurements for the individual observations of REJ2214-492, as a function of the average signal-to-noise in the 1050-1055Å range.

In the case of *FUSE* observations, there is an additional effect to consider, the location of the source in the aperture. While the alignment of one of the four spectrographs is accurately and reliably maintained through the attitude control system, there is flexure in the relative alignment of each of the other spectrographs due to thermal effects. While great care is taken to minimise this effect, source drift within the apertures cannot be entirely eliminated and is not taken into account in either the exposure time or flux calibration. In extreme cases, the source may drift completely out of the aperture. In this event the true exposure will be less than the expected time and the flux calibration will yield a clearly erroneously low value. However, if the source remains within the aperture for the complete duration of the observation, but is subject to vignetting, or only drifts out for a very short net period, the result will be less obvious. The "worm" effect, discussed in section 2.1 may also be sensitive to the precise source alignment. A comparison of the flux levels of the repeated observations of the same star (e.g. REJ2214-492, Fig. 8) that we have used, in which we have mostly excluded any gross alignment problems, shows scatter at the few tens of percent level, which might be an indication of the behaviour discussed above. Indeed, Barstow et al. (2001) have also demonstrated that such modest errors in the flux calibration can give rise to the systematic errors in $T_{\rm eff}$ and log *g* that we report here. However, the overall degree of consistency between the results from the various observations that we have used is a good indication that our results are not unduly sensitive to source alignment problems and the effects of the "worm". Note that the slight trend for apparently decreasing flux is erroneous as the observations are not time ordered, but follow the sequence of programme and observation numbers. There is one observation (number 13), which has a much lower flux than the others, indicating a misalignment/overestimate exposure. This is an unusual occurrence, particularly as it happened while using the LWRS, which should be less sensitive to alignment problems. An inspection of the housekeeping records shows that this observation was made during a period when the satellite had an unusual pointing problem, associated with a reaction wheel running at a speed that was a harmonic of one of the telescope resonant frequencies (J. Kruk, personal communication). This yielded a larger than normal pointing jitter. Interestingly, this overall effect seems to have just produced a simple scaling of the flux across the whole spectrum and does not lead to obviously anomalous values of $T_{\rm eff}$ or log *g*.

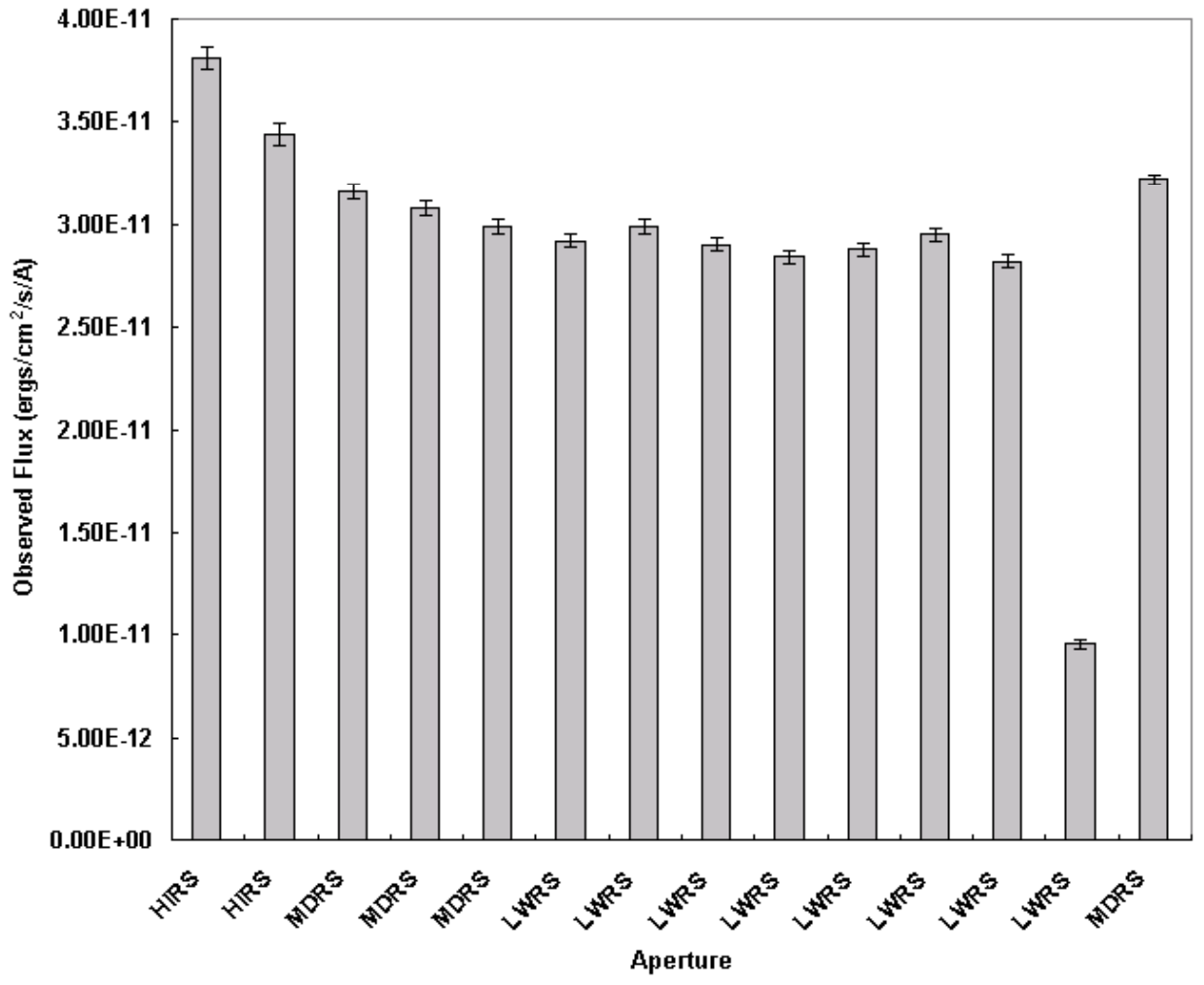

**Figure 8.** The measured average flux at in the 1050-1055Å range for each observation of REJ2214-492 with the associated 1σ statistical errors. The observations number run from 1 to 14, from left to right and each is labelled with the aperture used.

### 5.2 Comparison of Balmer and Lyman line measurements of $T_{\rm eff}$ and log *g*

Following the discussion of section 5.1, above, we are now able to make realistic estimates of the true observational errors inherent in the measurement of $T_{\rm eff}$ and log *g*, using the Balmer and Lyman lines. To make direct comparisons between each measurement technique we have plotted the simple mean values, where multiple observations exist, of the Lyman $T_{\rm eff}$ and log *g* against the Balmer $T_{\rm eff}$ and log *g* in Figs. 9 and 10 respectively. The error bars displayed are derived from the variance of the sequence of observations for each star, except where only one observation is available. Then we use just the statistical error, noting the comments in section 5.1, which indicate that these errors are at least a factor 2-3 smaller than the realistic values. Interestingly, the Lyman line measurements generally give smaller uncertainties in both $T_{\mathrm{eff}}$ and log $g$, even though the Balmer line spectra have a range of signal-to-noise (~50-100) that is very similar to the Lyman line data. This indicates that, for stars in the temperature range covered by this sample, the Lyman series provides a more sensitive determination of $T_{\mathrm{eff}}$ and log $g$ than does the Balmer series.

### *5.2.1 Effective temperature comparisons*

At temperatures below ~50000K, there is very good agreement between the values of $T_{\mathrm{eff}}$ determined from the Lyman and Balmer line observations (Fig. 9), within the measurement uncertainties. However, at higher temperatures, the Lyman measurements give consistently higher values than the Balmer results, with the exception of PG1342+444.

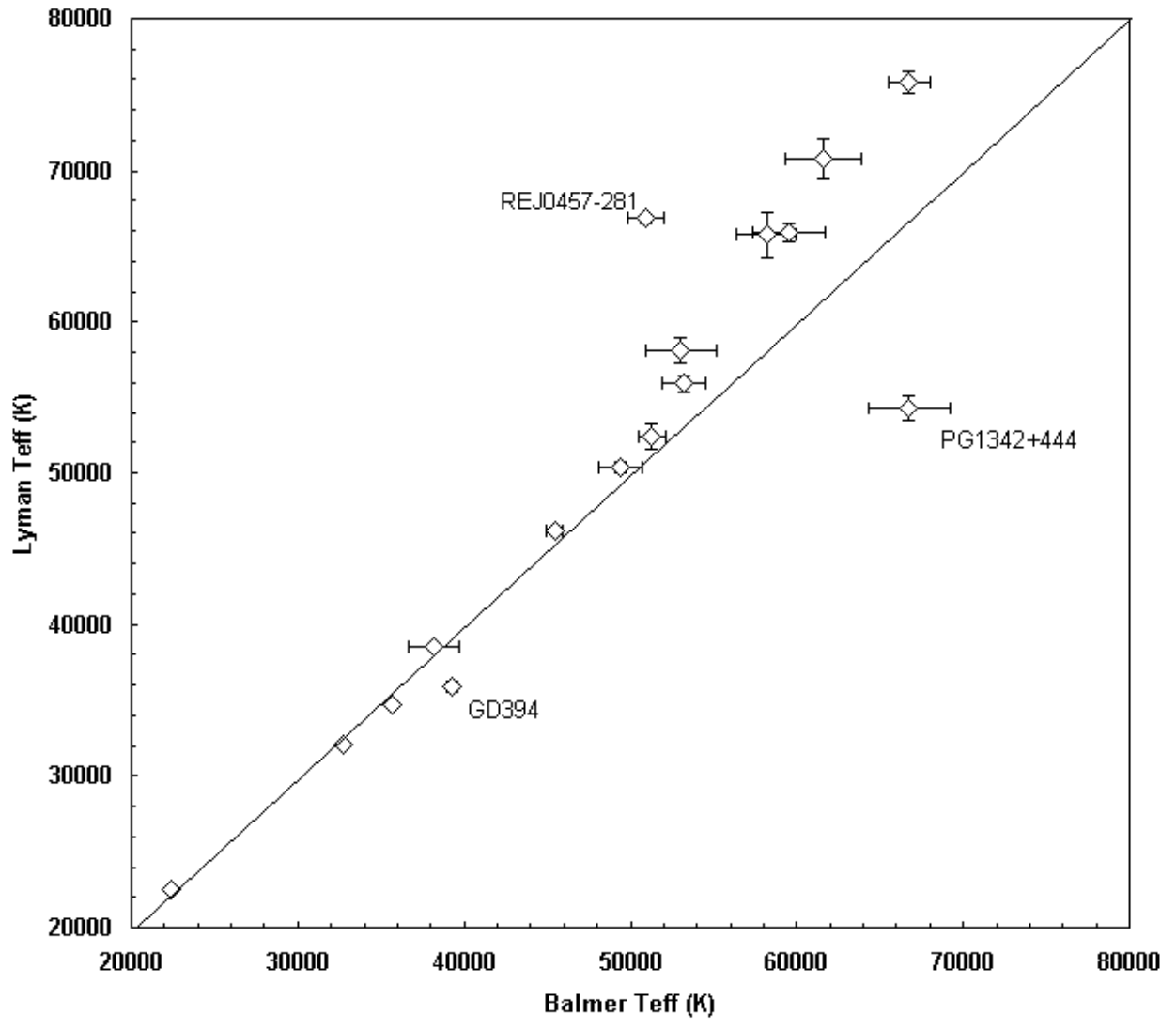

**Figure 9.** Scatter plot of the simple mean values of $T_{\mathrm{eff}}$ measured using the ground-based Balmer and FUSE Lyman lines. The error bars are calculated from the variance of the values in multiple observations or are the statistical 1σ error for single observations. The solid line corresponds to equal Balmer and Lyman line temperatures.

Among the hottest group of stars PG1342+444 and REJ0457-281, labelled in Fig, 9, stand out from the rest of the sample by departing significantly from the observed Lyman vs. Balmer line trend. Uniquely, as already reported by Barstow et al. (2001, 2002), PG1342+444 has a Lyman line determined temperature that is significantly lower than the value measured from the Balmer lines. Conversely, the Lyman temperature of REJ0457-281 is significantly in excess of the Balmer line result, at a level several times that observed in the other stars in the sample. This same discrepancy was also observed in the earlier *ORFEUS* observation of REJ0457-281, reanalysed by Barstow et al. (2001), showing a consistent result for observations made with different instruments. With a clear trend observed, as described above, for all the other stars in the sample, it appears that these two objects are peculiar in some still to be identified way. Observations of more stars in this temperature range will be needed to establish just how unique these stars are. With a comparatively small sample, it remains possible that we are seeing a couple of outliers from a distribution around the general trend.

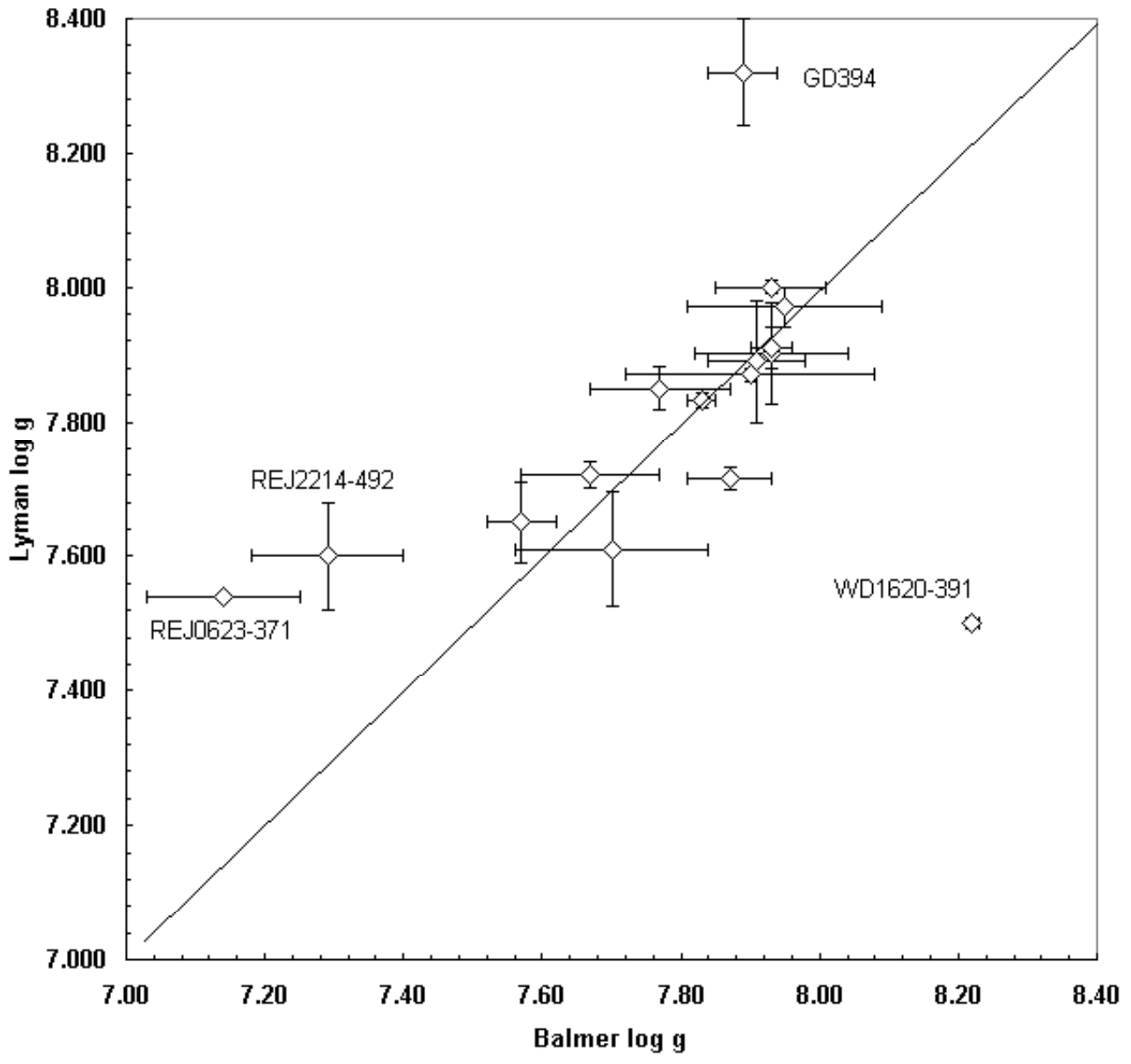

**Figure 10.** Scatter plot of the simple mean values of log $g$ measured using the ground-based Balmer and FUSE Lyman lines. The error bars are calculated from the variance of the values in multiple observations or are the statistical 1σ error for single observations. The solid line corresponds to equal Balmer and Lyman line gravity.

### *5.2.2 Surface gravity comparisons*

The picture is less clear-cut for the Lyman and Balmer log $g$ measurements (Fig. 10). In general there is good agreement between the measurements for most stars, but no particular trend is seen for the four stars that have differing Lyman and Balmer log $g$ values. Of these, GD394 may be peculiar. The gravity measured using the Lyman lines is significant larger than the Balmer result. However, Dupuis et al. (2000) have found GD394 to be photometrically variable in the EUV and have observed an extreme abundance of Si in the optical. Episodic accretion is proposed as the origin of a large EUV-dark spot on the surface, which may also explain the anomalous Si abundance. The presence of this feature, if associated with a different temperature to the rest of the star may affect the line profiles, compare to the homogeneous pure H atmosphere predictions. In addition any effect will be averaged in some way over the observation duration according to the phase of the 1.150d cycle. The phase coverage is unlikely to be the same for optical and far UV exposures. Although the discrepancy is less obvious in Fig. 9, the Balmer temperature is somewhat higher than the Lyman value.

In the case of WD1620-391, the Lyman determined gravity is lower than that obtained from the Balmer lines. Examination of the spectrum in Fig. 1 reveals the presence of quasi-molecular Lyman satellite lines, distorting the normal profiles (e.g. Wolff et al. 2001). Since, we did not include a treatment of these features in the stellar model atmosphere calculations, it is likely that the Lyman gravity is incorrect.

The two remaining stars, REJ2214-492 and REJ0623-371, are the two lowest gravity objects in the sample. If we exclude GD394 and WD1620-391 from consideration, they may define a trend of departure from the line of equal Lyman and Balmer gravity at low gravity, the former having the higher value, similar to the disagreement between Lyman and Balmer line temperatures. However, if real, this trend is less clear-cut than for the temperature comparison. It is notable that these objects are also among the hottest in the sample with significant quantities of heavy elements and strong absorption lines in the *FUSE* wavelength range. Apart from REJ1738+665, where there are a large number of interstellar $H_2$ absorption lines, we have had to remove more absorption lines from the *FUSE* spectra of these stars than any others. To test how much influence this has on the results of the spectral analysis, we prepared a sub grid of synthetic spectra, with heavy element spectral lines included, for these two stars. Within the statistical errors, the results are unaltered by the analysis with the more detailed models.

### 5.3 Implications for the model atmosphere calculations

We have examined the scope of the possible systematic errors in the analyses that have been performed and taken their magnitude into account in the comparisons of Lyman and Balmer observations shown in Figs. 9 and 10 and discussed above. In this context, it is clear that the observed differences are significant. If our understanding of the stars and their atmospheres was complete then we should have good agreement. Hence the discrepancies we report are an indication that the model calculations are deficient in some way for the heavy element-rich hottest DAs. In terms of the treatment of photospheric heavy elements, these models are highly advanced, dealing with many millions of transitions. However, the atmospheres are assumed to be homogeneous when there is contradictory evidence from the work of Holberg et al. (1999), Barstow, Hubeny & Holberg (1999), Dreizler & Wolff (1999) and Schuh, Dreizler & Wolff (2002). A self-consistent treatment with radiative levitation included in the model atmosphere calculations needs to be developed.

Although the model stratification is a likely explanation of the problem, there are other possibilities that should be considered in future model developments. The treatment of the hydrogen lines is well advanced in the detailed inclusion of the higher level lines and the line broadening theory, but most of the work in validating the input physics has been based on the Balmer line analyses, with little opportunity to consider the Lyman lines in detail. Treatment of the Lyman lines in the models may need further examination. However, the good agreement of the Balmer and Lyman line measurements at temperatures below 50000K suggests that the line broadening theory is not the problem.

The models computed for this analysis include most but not all of the heavy elements detected in the atmospheres of these stars. Al, P and S have been detected in the hottest stars. Although these elements have many fewer lines than Fe and Ni, which dominate the atmospheric structure, along with the strong resonance lines of C, N, O and Si, their inclusion may yield subtle modifications to the Lyman line profiles that can explain our observations.

In some of our earlier work, we have established that the values of $T_{\rm eff}$ and log $g$ determined from Balmer line analyses are sensitive to assumptions about the white dwarf photospheric composition. We have included heavy elements in the models used here at the known abundances. However, it is worth examining whether or not altering these abundances might have some influence on the Lyman/Balmer temperature discrepancy that we have observed. To do this we computed two further grids of models with 0.1 times and 10 times the nominal abundances and carried out a test on one *FUSE* observation from each of two stars, G191-B2B (obs 1) and REJ2214-492 (obs 7). We find that the heavy element content changes the magnitude of the difference between Lyman and Balmer values of $T_{\rm eff}$. The difference is larger for the 0.1 times abundance models and smaller when the abundance is 10 times the nominal value. However, the changes in the temperature difference are only ~20-30% and the discrepancy remains significant. We conclude that the observed temperature discrepancy is not very sensitive to assumptions about the photospheric abundances. It important to note that there is little scope for errors in the abundance determinations as large as the factors examined here (see Barstow et al. 1993). However, the fact that we do see some effect may be a hint that depth dependence of the heavy element abundances could be important.

It is important to remember that the Balmer and Lyman line measurements give good agreement for those stars with temperatures below 50000K and most surface gravities. Therefore, any stars that fall into this temperature range can be studied reliably using just the Lyman lines when the Balmer lines are not accessible. The results presented here also provide an empirical relationship between the Lyman and Balmer line derived temperatures above 50000K. Apart from exceptions like REJ0457-281and PG1342+444, the relationship appears to be single valued. Of course, until we have discovered what makes these two stars unusual, we cannot know, a-priori, which of the hot stars we observe do follow the relation and which do not.

All the stars with values of $T_{\rm eff}$ below 50,000K have more or less pure H atmospheres and were, therefore, analysed using pure H models. This included GD394, which is known to have some heavy element content. However, the heavy element grid calculations did not extend down to such a low temperature. It is interesting that the Lyman/Balmer temperature discrepancy is only associated with stars analysed using the heavy element blanketed models. To ascertain whether or not the temperature problem is particularly associated with the heavy element models, we repeated the analysis for selected spectra from those stars with $T_{\rm eff}$ above 50000K with pure H models. Although, as expected from the study of Barstow et al. (1998), the actual values of $T_{\rm eff}$ were higher than for the heavy element models, the difference between the Lyman and Balmer line results is of a similar magnitude.

Since the Balmer lines have usually been used to measure $T_{\rm eff}$ and log $g$ for the hot DA white dwarfs, it is tempting to assume that the Balmer line analyses give the "correct" values for the hottest white dwarfs. This is not necessarily the case. Whatever the underlying physical explanation of the high temperature discrepancy, the effect could modify both series of absorption lines, with the true value of $T_{\rm eff}$ between the values we determine. It might even be the case that the Lyman lines give the more reliable temperature determination.

## 6 CONCLUSION

With the availability of the *FUSE* data archive and observations from our own Guest Observer programmes, we examined the use of the Lyman series to determine the values of $T_{\rm eff}$ and log $g$ for a sample of 16 hot white dwarfs. Having a source of data produced by a single instrument and processed with a uniform pipeline, we were able to eliminate some of the possible systematic differences between observations of the same or different stars associated with different instruments. However, it is clear from this study that systematic errors in the overall observation, data reduction and analysis procedures dominate the measured uncertainties. We have used the scatter in values derived from multiple observations of some stars to determine more realistic errors in the measurements than obtained just from the statistical error values. The new results partially reproduce our earlier study, where we studied a more limited stellar sample with data from *HUT*, *ORFEUS* and a few early *FUSE* observations, showing that Balmer and Lyman line determined temperatures are in good agreement up to ~50000K. However, above this value there is an increasing systematic difference between the Lyman and Balmer line result, the former yielding the higher temperature. At the moment, there is no clear explanation of this effect but we think that it is most likely associated with deficiencies in the detailed physics incorporated into the stellar model atmosphere calculations. Even so, the data do demonstrate that, for temperatures below 50000K, the Lyman lines give reliable results. Furthermore, for the hotter stars, a useful empirical calibration of the relationship between the Lyman and Balmer measurements has been obtained, that can be applied to other *FUSE* observations.

## ACKNOWLEDGEMENTS

The work reported in this paper was based on observations made with the NASA-CNES-CSA *FUSE* observatory. The Johns Hopkins University operates *FUSE* for NASA, under contract NAS5 32985. We are indebted to Jeff Kruk of the *FUSE* operations team for his assistance in the provision of useful information on the various calibration and data quality issues discussed in this paper. MAB, MRB, SAG and ANL were supported by PPARC, UK. JBH wishes to acknowledge support from NASA grant NAG5 9181.